\documentclass{elsart}
\usepackage{graphicx}
\begin{document}
\begin{frontmatter}
\title{Using magnetic levitation to produce cryogenic targets for inertial fusion energy: experiment and theory}
\author{D. Chatain}
\and
\author{V. S. Nikolayev\thanksref{ghi}}
\address{CEA/DSM/SBT/ESEME, CEA Grenoble,
17, rue des Martyrs, 38054, Grenoble Cedex 9, France}
\thanks[ghi]{Mailing address: ESEME-CEA, Institut de Chimie de la Mati\`ere Condens\'{e}e de Bordeaux, CNRS,
Avenue du Dr. Schweitzer, 33608 Pessac Cedex, France; e-mail: \texttt{vnikolayev@cea.fr}}
\date\today
\begin{abstract}
We present experimental and theoretical studies of magnetic levitation of hydrogen gas bubble surrounded
by liquid hydrogen confined in a semi-transparent spherical shell of 3 mm internal diameter. Such shells
are used as targets for the Inertial Confinement Fusion (ICF), for which a homogeneous (within a few
per-cent) layer of a hydrogen isotope should be deposited on the internal walls of the shells. The gravity
does not allow the hydrogen layer thickness to be homogeneous. To compensate this gravity effect, we have
used a non-homogeneous magnetic field created by a 10~T superconductive solenoid. Our experiments show that
the magnetic levitation homogenizes the thickness of liquid hydrogen layer. However, the variation of the
layer thickness is very difficult to measure experimentally. Our theoretical model allows the exact shape
of the layer to be predicted. The model takes into account the surface tension, gravity, van der Waals,
and magnetic forces. The numerical calculation shows that the homogeneity of the layer thickness is
satisfactory for the ICF purposes. \end{abstract}
\end{frontmatter}

\section{Introduction}

Several concepts have been proposed for the design of a commercial power plant based on Inertial Fusion
Energy production \cite{1,2}. Targets are direct or indirect drive targets but must be at cryogenic
temperature \cite{3}. They must be injected in the vacuum chamber of the reactor at a rate of about 5 Hz
and a speed depending on the temperature and the residual pressure of the vacuum vessel \cite{4}.  The
targets are then tracked and hit on-the-fly with laser or heavy ion beams \cite{5}.

The targets are hollow spherical shells made of beryllium or polystyrene. Their diameter ranges from 2 to
5 mm. Their internal wall must be covered with a solid layer of deuterium or a mixture of deuterium and
tritium of several hundred microns in thickness. The thickness of the layer must be uniform within a few
percent. If tritium is present in the mixture, the beta energy produced by tritium naturally drives the
solid to a uniform thickness layer covering the internal walls of the shell \cite{6,7}. If tritium is not
present in the liquid layer, it stays at the bottom because of gravity. In this paper, we describe how, by
using the diamagnetic properties of the hydrogen, we can compensate this gravity effect and obtain a
homogenous thickness of liquid layer inside the sphere before freezing it.

\section{Forces that act on hydrogen molecules}

In order to obtain the homogeneous thickness of the liquid layer
on the inner walls of a hollow spherical shell, one needs to
satisfy simultaneously two conditions:
\begin{itemize}
  \item the shape of the gas bubble inside the liquid should be spherical,
  \item the gas bubble should levitate in the middle of the shell.
\end{itemize}
We show now how the various forces influence the satisfaction of these conditions.

\subsection{Surface tension}

The force of the surface tension tends to minimize the interface area. Therefore, the surface tension
helps to maintain the spherical shape of the gas bubble. Obviously, we need to look for the conditions
where the value $\sigma$ of the surface tension is large. As a matter of fact, $\sigma$ is the decreasing
function of the temperature $T$ and goes to zero at the critical temperature $T_c$, which is about 33 K
for hydrogen. The working temperature should thus be at least several degrees smaller than $T_c$.

The contribution of the surface tension (i.e. the Laplace
pressure) is inversely proportional to the radius of curvature of
the interface. For the thicker liquid layer the radius of the gas
bubble is smaller and the Laplace pressure is thus larger.
Therefore, the thickness of the thicker liquid layer is {\em a
priori} more homogeneous (provided that the gas bubble is
levitated in the middle of the shell) than the thickness of the
thinner liquid layer. As a consequence, we need to analyze the
homogeneity only for the thinnest layer under consideration, which
is 200~$\mu$m. If for this case the homogeneity 1\% criterion is
satisfied, it will be satisfied for all larger thicknesses.

\subsection{Van der Waals forces}

Since the hydrogen completely wets the solid shell (zero contact
angle), the van der Waals force manifests itself as an attraction
between the hydrogen molecules and the solid wall \cite{Kim}. It
thus tends to create the layer of the densest (liquid) phase at
the shell wall, leaving the less dense phase (gas) in the middle
of the shell. However, the value of this force is very small in
comparison with the surface tension. While the van der Waals force
influences strongly \cite{Kim} the shape of the layers of
microscopic (of the order of 1 $\mu$m) thickness, we do not expect
a strong effect for the case of the thick liquid layers (average
thickness larger than 200 $\mu$m), which we analyze in this
report.

We carry out all our calculations for the non-retarded van der Waals interactions instead of the more
suitable (because of the large layer thickness) retarded expression. The reason is that the retarded
interactions are weaker \cite{Kim} and would result in even smaller contribution.

\subsection{Gravity and magnetic forces: magnetic levitation}

The gravitational force per unit volume $f_g=\rho g$ is proportional to the mass density $\rho$ of
hydrogen, $g$ being the gravitational acceleration. Since the liquid mass density $\rho_L$ is larger than
the gas mass density $\rho_G$, the resulting Archimedes force $\Delta \vec{f}_g=(\rho_G-\rho_L)\vec{g}$
that acts on the gas bubble tends to increase the thickness of the liquid layer on the bottom of the shell
by lifting the bubble upwards. Gravitation thus needs to be compensated.

The gravity compensation by means of the static magnetic field is based on the expression for the magnetic
force per unit volume $\vec{f}_m=\chi\,\nabla(B^2)/2\mu_0$ that acts on a body with the magnetic
susceptibility $\chi$, where $B$ is the magnetic induction that would be created by the same solenoid in
the free space and $\mu_0$ is the magnetic permeability of free space. The magnetic susceptibility is
proportional to the mass density $\rho$, $\chi/\rho=\alpha$ (see Table~\ref{tab2}).
\begin{table}[htb]
\caption{Parameters of hydrogen at $T= 20$ K \cite{11}.}\label{tab2}
\begin{tabular}{|c|c|c|c|}\hline
 Description & Notation & Value & Units \\ \hline
  Magnetic susceptibility/mass density & $\alpha$ & $-2.51\cdot 10^{-8}$& m$^3$/kg \\
  Surface tension &  $\sigma$ & 0.002 & N/m\\
  Mass density of liquid phase &  $\rho_L$ & 71.41 &  kg/m$^3$\\
  Mass density of gas phase &  $\rho_G$ & 1.19 &  kg/m$^3$\\
 \hline
\end{tabular}
\end{table}
The magnetic force that acts on the bubble is thus (see Appendix)
\begin{equation}\label{dfm}
  \Delta \vec{f}_m=(\rho_G-\rho_L)\alpha\,\nabla(B^2)/2\mu_0
\end{equation}
and the condition of the bubble levitation is $\Delta \vec{f}=\Delta \vec{f}_m+\Delta \vec{f}_g=0$.

\begin{figure}[htb]
  \begin{center}
  \includegraphics[height=5cm]{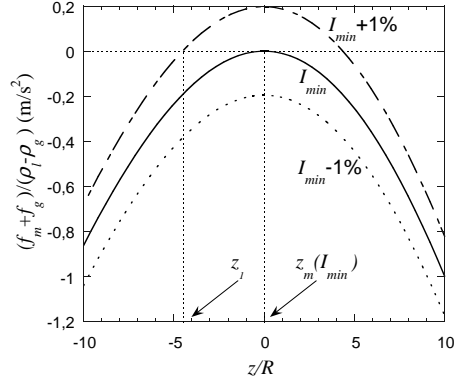}
  \end{center}
  \caption{The effective force per unit mass that acts on a small gas bubble in presence of the magnetic
  field of the solenoid versus $z$-coordinate along the solenoid axis calculated for tree values of the
  solenoid current: $I_{min}$, $I_{min}\cdot 1.01$, and $I_{min}\cdot 0.99$.}\label{force}
\end{figure}

Note that $\nabla(B^2)/2=B_z\, \mathrm{d}B_z\mathrm{d}z$ along the solenoid axis (axis $z$) and the curve
$\Delta \vec{f}(z)$ exhibits a maximum (see Fig.~\ref{force}) at some value of $z=z_m(I)$ that is almost
independent of the solenoid current $I$.

All magnetic field calculations reported in this article were performed with the code BOBOZ translated to
the C programming language. As an additional input to this code, the solenoid current in amperes
multiplied by the number of coils is required.

Exact gravity compensation \begin{equation}\label{equil}
  g={|\alpha|\over \mu_0} B_z\,{\mathrm{d}B_z\over\mathrm{d}z}
\end{equation}
can be achieved when $I\ge I_{min}$, where $I_{min}$ is a minimum current at which the exact compensation
is possible at all. In the experiments (see sec. \ref{exp}), the current $I_{min}$ is 60~A. In our
calculations we choose the point $z_m(I_{min})$ as a zero reference point. Its position is 8.502~cm above
the solenoid center. According to our calculations, $I_{min}$ corresponds to $1.8300\cdot10^6$
ampere-coils. The unknown (the solenoid documentation is not available) number of coils is obtained by
division of this number by $I_{min}$.

\begin{figure}[htb]
  \begin{center}
  \includegraphics[height=5cm]{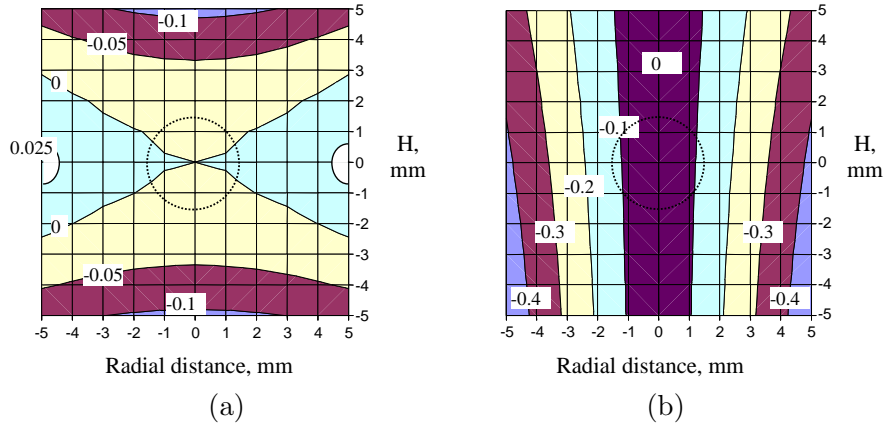}
  \end{center}
\hfil\small{(a)}\hfil\small{(b)}\hfil\caption{The iso-acceleration curves for (a) $\gamma_z$ and (b)
$\gamma_r$. The position of the sphere is indicated by the dotted lines. The acceleration values are given
in m~s$^{-2}$.}\label{field}
\end{figure}

Fig.~\ref{force} shows that when $I<I_{min}$, no compensation is possible at all, the force is non-zero
everywhere. When $I>I_{min}$, there are two points of compensation $z_1$ and $z_2$ such as $z_1<z_m<z_2$.
However, it can be shown that levitation in $z_2$ is not stable, so that the compensation can be achieved
in only one point $z_1<z_m$. Everywhere else, the hydrogen molecules are exposed to the residual
acceleration $\vec{\gamma}=|\alpha|/(2\mu_0)\;\nabla B^2-\vec{g}$ the vertical ($z$) and radial ($r$)
components of which are shown in Fig.~\ref{field}.
As we can see on these graphs, the radial component of the acceleration in the vicinity of the center of
the sphere is more important than the axial component. Inside a 3~mm diameter sphere, the maximum radial
acceleration is about 0.125~m~s$^{-2}$, while the vertical acceleration is ten times lower. The radial
acceleration is directed toward the vertical axis. This means that the interface can deviate from the
spherical shape because it follows the variation of $\vec{\gamma}$. It is thus important to know if the
homogeneous thickness of the liquid layer can in principle be achieved with the magnetic field created by
the superconductive solenoid.

\section{Experimental}\label{exp}

\subsection{Test facility}

\begin{figure}[hbt]
 \begin{minipage}[b]{0.5\textwidth}
  \begin{center}
  \includegraphics[height=6cm]{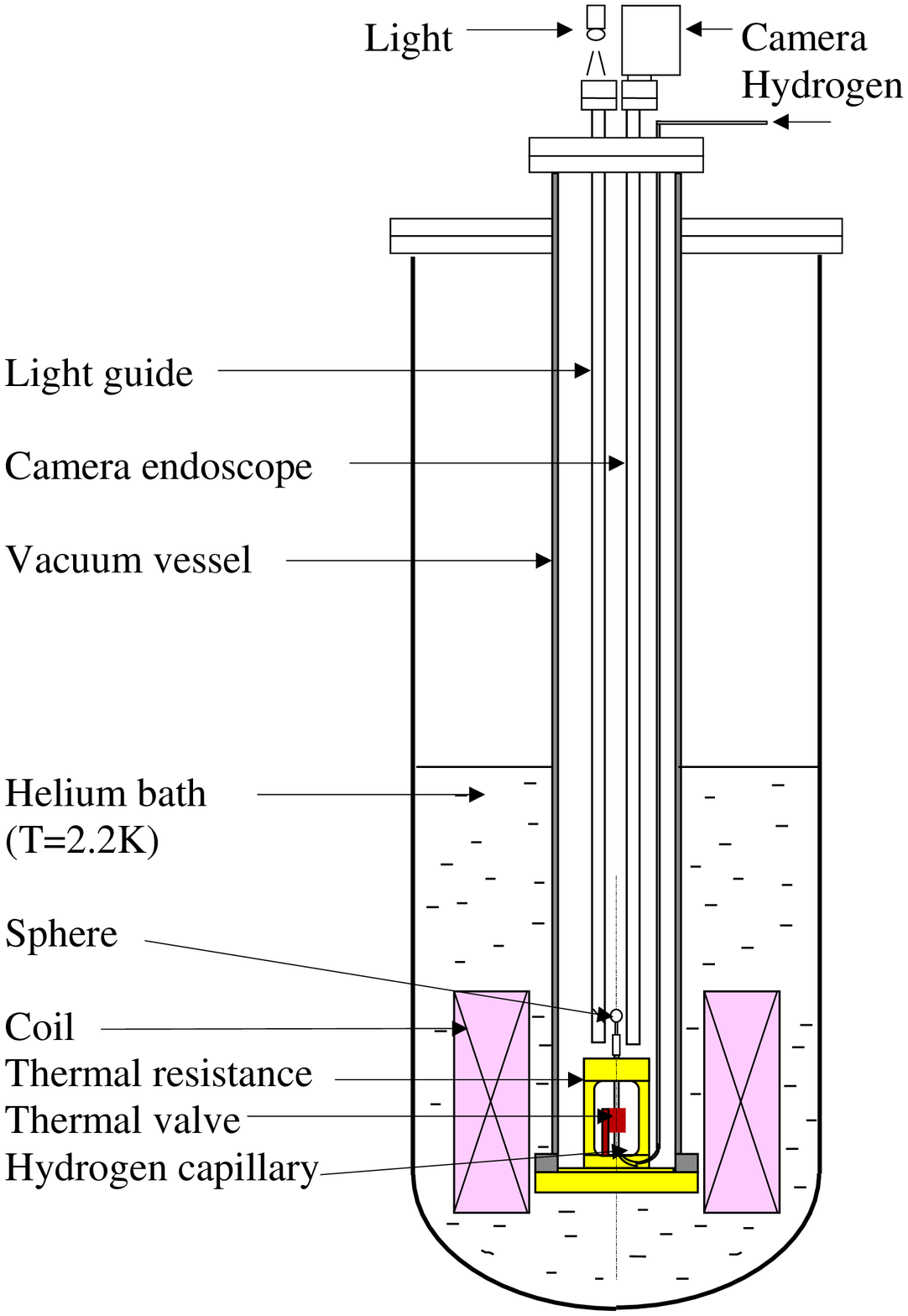}\\
  \small{(a)}
  \end{center}
  \end{minipage}%
  \begin{minipage}[b]{0.5\textwidth}
  \begin{center}
  \includegraphics[height=6cm]{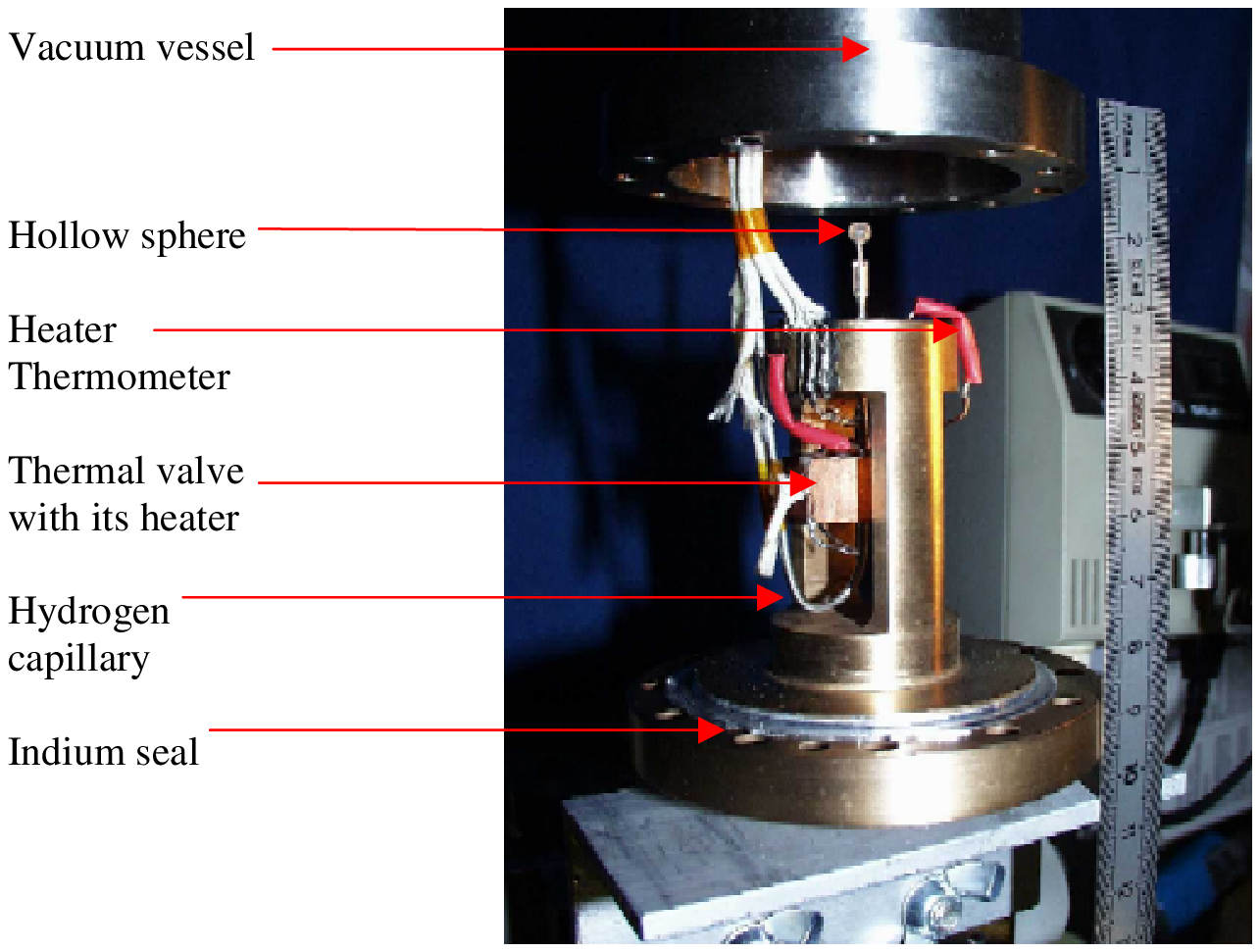}\\
  \small{(b)}
  \end{center}
  \end{minipage}
\caption{(a) A scheme of the cryostat. (b) A photograph of the bottom of the cryostat showing the
sphere.}\label{cryo}
\end{figure}

The test facility (Figs.~\ref{cryo}) consists of a superconductive solenoid immersed into a helium bath.
The vacuum vessel containing the sphere is introduced into the core of the solenoid. The sphere is
illuminated from the top of the cryostat with a light guide. Observation is performed with an endoscope
and a CCD camera. The bottom of the cryostat with the sphere is shown in Fig.~\ref{cryo}b. The sphere
center is placed at 15~mm from the top of the solenoid i.e. into the calculated point $z_m(I_{min})$. The
hydrogen is introduced into the sphere by capillary. The characteristics of the solenoid are listed in
Table~2.
\begin{table}[ht]
\caption{Parameters of the superconductive solenoid.}
\begin{tabular}{|c|c|c|}\hline
 Description & Value & Units \\ \hline
  Inner radius & 48 & mm \\
   Outer radius & 93 & mm \\
   Total  height & 20 & cm \\
   $B$ at 4.2 K and $I=53$ A & 8 & T\\
   $B$ at 2.17 K and $I=67$ A & 10 & T\\
   Critical current at 2.17 K & 72 & A\\
 \hline
\end{tabular}
\end{table}

The following operations are performed in order to condense and levitate the hydrogen in the sphere:
\begin{itemize}
\item The vacuum vessel is pumped out to about $10^{-6}$~mb and the capillary is pumped out to about 0.1~mb.
\item The solenoid and the
vacuum vessel are cooled down to 2.2~K. \item The thermal valve is heated (a power of 300~mW is necessary)
\item The sphere is heated to about 20~K. \item Hydrogen is slowly introduced into the capillary.
\item When a sufficient quantity is condensed inside the sphere, the heater of the valve is cut so that
an ice plug clogs up the capillary. \item The current in the solenoid is increased to about 60~A to
compensate the gravity. \item The temperature of the sphere can be decreased below the triple point (13~K)
to freeze the liquid.
\end{itemize}

\subsection{Sphere}

\begin{figure}[tb]
 \begin{minipage}[b]{0.5\textwidth}
  \begin{center}
  \includegraphics[height=6cm]{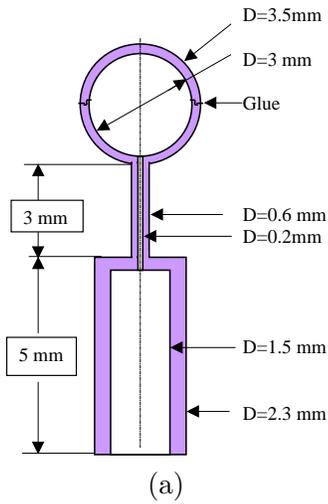}\\
  \small{(a)}
  \end{center}
  \end{minipage}%
  \begin{minipage}[b]{0.5\textwidth}
  \begin{center}
  \includegraphics[height=6cm]{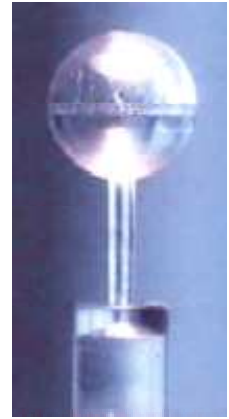}\\
  \small{(b)}
  \end{center}
  \end{minipage}
\caption{(a) Design of the transparent hollow sphere. (b) A photograph of the sphere.}\label{sphere}
\end{figure}

The sphere was made of Plexiglas (PMMA) machined in two hemispheres according to the design shown in
Figs.~\ref{sphere}. The two hemispheres were glued with \emph{Epoxy 501} stick.

\subsection{Effect of the magnetic field}

\begin{figure}[htb]
  \begin{center}
  \includegraphics[width=\textwidth]{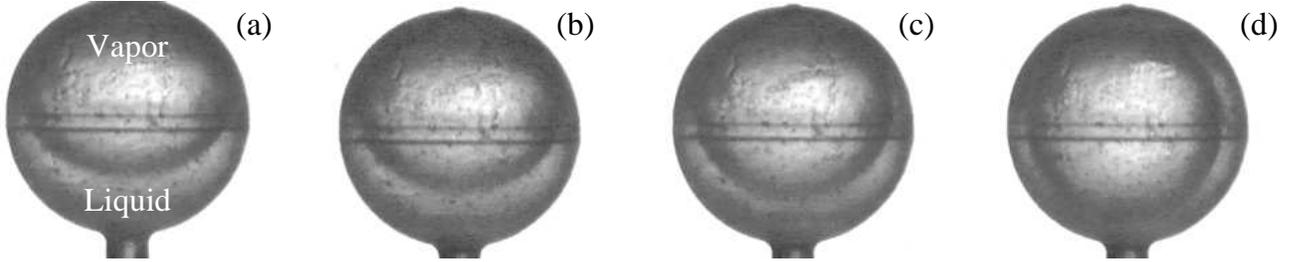}
  \end{center}
\caption{The sphere half filled with liquid hydrogen at a temperature close to the triple point (14 K)for
the different values of the current $I$ in the solenoid that correspond to the different amplitudes of the
magnetic field: (a) $I=0$, (b) $I=30$ A, (c) $I=50$ A, (d) $I=60$ A. The gravity is completely compensated
by the magnetic force. The gas bubble is stable and well centered.}\label{Sph_field}
\end{figure}

As one can see in Figs.~\ref{Sph_field}, it was possible to condense the hydrogen in the sphere. The
optical imperfections of the sphere did not allow a high image quality to be obtained. Consequently, a
precise measurement of the homogeneity of the layer thickness was not possible. Nevertheless, we can see
in Fig.~\ref{Sph_field}d that the gas-liquid interface deformation due to the residual gravity force does
not seem to be large.

The figures below show the effect of the magnetic field on the shape of the vapor bubble.

\subsection{Effect of the liquid/vapor volume ratio}

\begin{figure}[htb]
  \begin{center}
  \includegraphics[width=\textwidth]{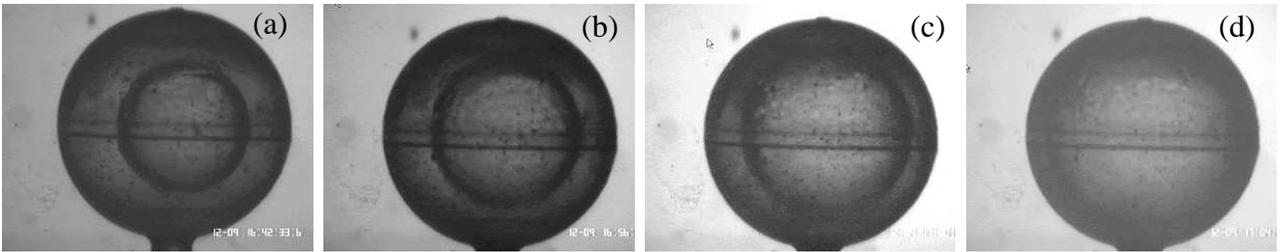}
  \end{center}
\caption{Same as in Fig.~\ref{Sph_field}d for the gravity compensation field and for the different liquid
layer thickness: (a) 520 $\mu$m, (b) 350 $\mu$m, (c) 200 $\mu$m, (d) It becomes difficult to measure the
liquid layer thickness. }\label{Sph_thick}
\end{figure}

The images in Figs.~\ref{Sph_thick} show the effect of the liquid/gas volume ratio for a gravity completely
compensated by the magnetic field. This ratio controls the thickness of the liquid layer.

\section{Numerical modeling}

\subsection{Mathematical formulation}\label{sec3}

The equilibrium shape of the interface can be found by two different approaches. One of them consists in
direct numerical minimization of the free energy of the system. In this work we adopt another, variational
approach, in which the minimization itself is performed analytically. It results in a variational equation
that should be solved numerically to obtain the interface shape. The general form of this variational
equation is the Laplace equation
\begin{equation}\label{laplace}
  K\sigma=\Delta p,
\end{equation}
where $K$ is the local curvature of the interface, and $\Delta p$ is the difference between the forces per
unit area that act on the interface from the liquid and gas sides. There are several contributions to
$\Delta p$\/:
\begin{equation}\label{dp}
  \Delta p=\Delta p_g+\Delta p_m+\Delta p_w+\lambda
\end{equation}
that correspond to gravitation ($\Delta p_g$), magnetic field ($\Delta p_m$), and the van der Waals force
($\Delta p_w$). The constant $\lambda$ is a Lagrange multiplier that appears as a result of the constrained
gas volume $V_G$. $\lambda$ can be viewed otherwise as an unknown {\it a priori} difference of pressures
inside the liquid and gas phases.

The mathematical expression for the curvature $K$ depends on the choice of the reference system and the
independent variable. It is convenient to use the cylindrical $(r,z)$ coordinate system because the
solenoid ($z$) axis is vertical and $\Delta p=\Delta p\,(r,z)$ is thus cylindrically symmetric. In this
reference system, the interface is fully defined by its half-contour for which $r>0$.

None of the variables $r,z$ can be chosen as independent because for the closed interface contour (at
least) two values of $z$ exists for each value of $r$ and vise-versa. We choose as an independent variable
the curvilinear coordinate $l$ that varies along the interface contour counter-clockwise with $l=0$ at the
point on the symmetry axis where $r=0$. Using this parameterization, $z=z(l)$ and $r=r(l)$. Since $l$
measures the running length along the interface contour, the following equation is valid
\begin{equation}\label{sum}
  z'^2+r'^2=1,
\end{equation}
where prime means the derivative $\mathrm{d}/\mathrm{d}l$. The expression for the local curvature then
takes the form
\begin{equation}\label{K}
  K=r'z''-r''z'+z'/r.
\end{equation}
By introducing an auxiliary function $u(l)$, one can reduce (\ref{laplace}) (with $K$ given by (\ref{K}))
to the set of the first-order ordinary differential equations:
\begin{equation}\label{syst}
\left\{
  \begin{array}{l}
    u'=\Delta p\,(r,z)/\sigma -\sin u/r \\
    r'=\cos u \\
    z'=\sin u
  \end{array}
  \right.
\end{equation}
The physical meaning of the variable $u$ can be found out by dividing two last equations: $u$ is the angle
between the $r$ axis and the tangent to the interface contour.

The gas bubble volume $V_G$ is fixed. This condition allows $\lambda$ to be determined from the equation
\begin{equation}\label{voll}
  V_G=\pi\int\limits_0^L r^2 \sin u\;{\rm d}l,
\end{equation}
where $r=r(l)$ and $u=u(l)$ are the solutions of the set (\ref{syst}) and $L$ is the unknown {\it a priori}
half-length of the interface contour.

Four boundary conditions for the set (\ref{syst}) should be specified at the points $l=0$ and $l=L$. Three
of them serve to determine the integration constants in (\ref{syst}) and the fourth serves to determine
$L$. There are two possible types of the boundary conditions. The first type corresponds to the case of the
continuous liquid layer \cite{Kim} and reflects the symmetry of the contour:
\begin{equation}
\left\{
  \begin{array}{l}
    u(0)=0,\\
    r(0)=0,\\
    u(L)=\pi,\\
    r(L)=0.
  \end{array}
  \right.
\end{equation}
The boundary conditions of the second type should be specified when the liquid layer is discontinuous when
the van der Waals forces are neglected in the calculation and the point $l=L$ is the triple
(gas-liquid-wall) contact point. Since the contact angle is zero, the boundary conditions take the form
\begin{equation}
\left\{
  \begin{array}{l}
    u(0)=0,\\
    r(0)=0,\\
    r(L)=R\sin u(L),\\
    z(L)=-R\cos u(L),
  \end{array}
  \right.
\end{equation}
where $R$ is the internal radius of the shell.

The mathematical problem is now complete. However, it is difficult to solve because of the moving boundary
conditions specified at the unknown upper boundary $L$. This problem can be reduced to the simpler problem
with the fixed boundary conditions by the following mathematical trick. We introduce new independent
variable $\xi=l/L$ and two more dependent variables $L$ and $\lambda$. The set (\ref{syst}) then reduces to
\begin{equation}\label{set}
\left\{
  \begin{array}{l}
    u'=L\,\Delta p\,(r,z)-\sin u/r \\
    r'=L\,\cos u \\
    z'=L\,\sin u \\
    L'=0\\
    \lambda'=0
  \end{array}
  \right.
\end{equation}
where $u$,$r$,$z$,$L$,$\lambda$ are supposed to be the functions of $\xi$, prime means now the derivatives
$\mathrm{d}/\mathrm{d}\xi$, and from now on we express $r$, $z$, and $L$ in the units of $R$ and $\Delta
p$ in the units $\sigma/R$. Five unknown integration constants for these five equations should be found
from the condition
\begin{equation}\label{vol}
  V_G=\pi L\int\limits_0^1 r^2 \sin u\;{\rm d}\xi,
\end{equation}
and from the four fixed boundary conditions (specified at $\xi=0,1$) for the problem of the first type
(continuous liquid layer)
\begin{equation}\label{bound1}
\left\{
  \begin{array}{l}
    u(0)=0,\\
    r(0)=0,\\
    u(1)=\pi,\\
    r(1)=0,
  \end{array}
  \right.
\end{equation}
or of the second type (discontinuous liquid layer, i.e. direct wall-gas contact)
\begin{equation}\label{bound2}
\left\{
  \begin{array}{l}
    u(0)=0,\\
    r(0)=0,\\
    r(1)=\sin u(1),\\
    z(1)=-\cos u(1).
  \end{array}
  \right.
\end{equation}

The set of equations (\ref{dp},\ref{set},\ref{vol}) with the boundary conditions (\ref{bound1}) or
(\ref{bound2}) provide two fully defined fixed boundary mathematical problems if the functional forms of
$\Delta p_m\,(r,z)$, $\Delta p_g\,(r,z)$, and $\Delta p_w\,(r,z)$ are known. The magnetic induction $B$
that enters the first of them (\ref{dpmn}) is calculated using the code BOBOZ. The second is given by
$\Delta p_g\,(r,z)=\mathrm{Bo}\cdot z$, where Bo is the non-dimensional Bond number
\begin{equation}\label{bo}
  \mathrm{Bo}=(\rho_L-\rho_V)gR^2/\sigma.
\end{equation}

The expression for the van der Waals contribution was calculated in \cite{Kim}. For the non-retarded
interaction $$\Delta p_w\,(r,z)=C_w[R_e^3(R_e^2-d)^{-3}-(1-d)^{-3}],$$ where $d=r^2+z^2$, $R_e$ is the
external shell radius (1.75 mm) expressed in the units $R$ and
\begin{equation}\label{vdW}
  C_w={4\pi\over 3}(\rho_L-\rho_V){b_{HS} N_A^2\rho_S\over\sigma R^2 m_S m_H}
\end{equation}
is the non-dimensional number that reflects the strength of the van der Waals forces relative to the
surface tension, $b_{HS}\approx 4\cdot10^{-78}$ J\,m$^6$ is the London constant for the interaction of the
shell and hydrogen molecules, $N_A$ is the Avogadro number, $\rho_S$ and $m_S$ is the mass density and the
molecular weight of the shell material, and $m_H$ is the molecular weight of hydrogen.

\subsection{Calculation details}

The mathematical problem (\ref{dp}), (\ref{set}), (\ref{vol}), (\ref{bound2}) with the boundary conditions
of the second type was solved using the Shooting Method \cite{NR} for the two-point boundary value
problems. This algorithm was modified slightly to include the condition (\ref{vol}) into the function {\em
score} of \cite{NR}. The integral in (\ref{vol}) was calculated using the Simpson method. The set of the
ordinary differential equations (\ref{set}) was solved using the {\em rkdumb} function of \cite{NR} with
100 steps, which was sufficient to achieve the accuracy of 10$^{-6}$. The mathematical problem (\ref{dp}),
(\ref{set}), (\ref{vol}), (\ref{bound1}) with the boundary conditions of the first type should be solved
using the Shooting to the Fitting Point Method \cite{NR} because of the singularity $r=0$ at the boundary
point $\xi=1$.

The calculations were performed for the values of the parameters shown in Table 2. The gas bubble volume
$V_G$ is defined by the average value for the liquid layer thickness of 200~$\mu$m.

\subsection{Levitation at minimum compensation current}

\begin{figure}[htb]
  \begin{center}
  \includegraphics[height=5cm]{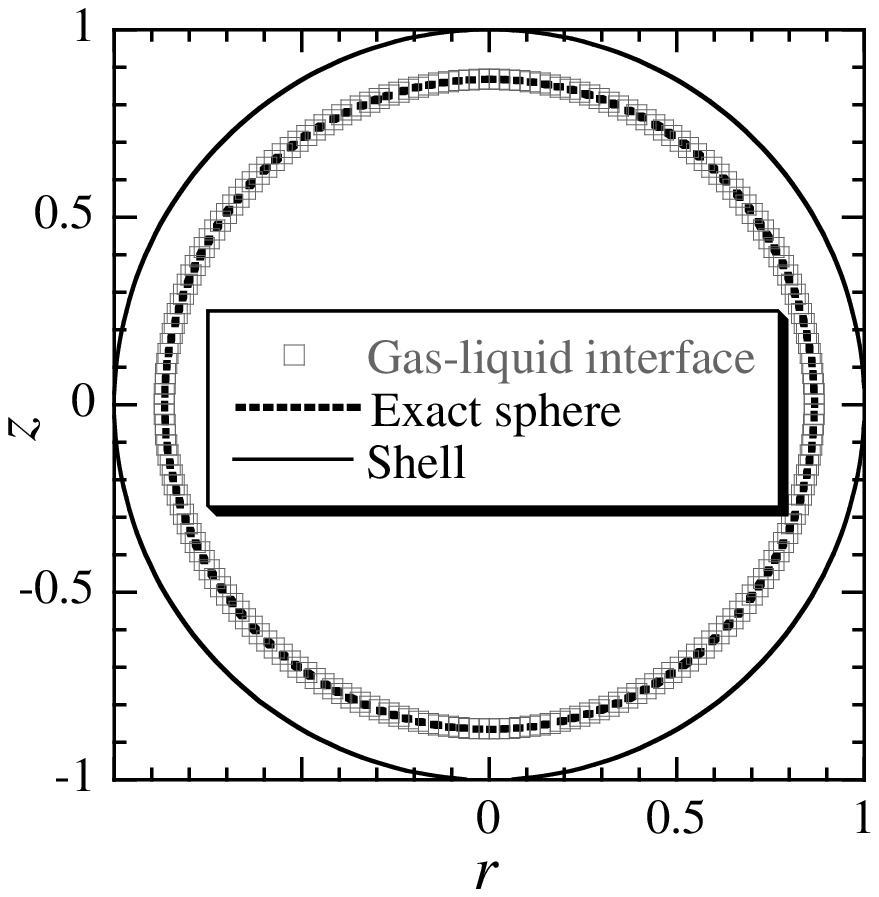}\hspace*{7mm}\includegraphics[height=5cm]{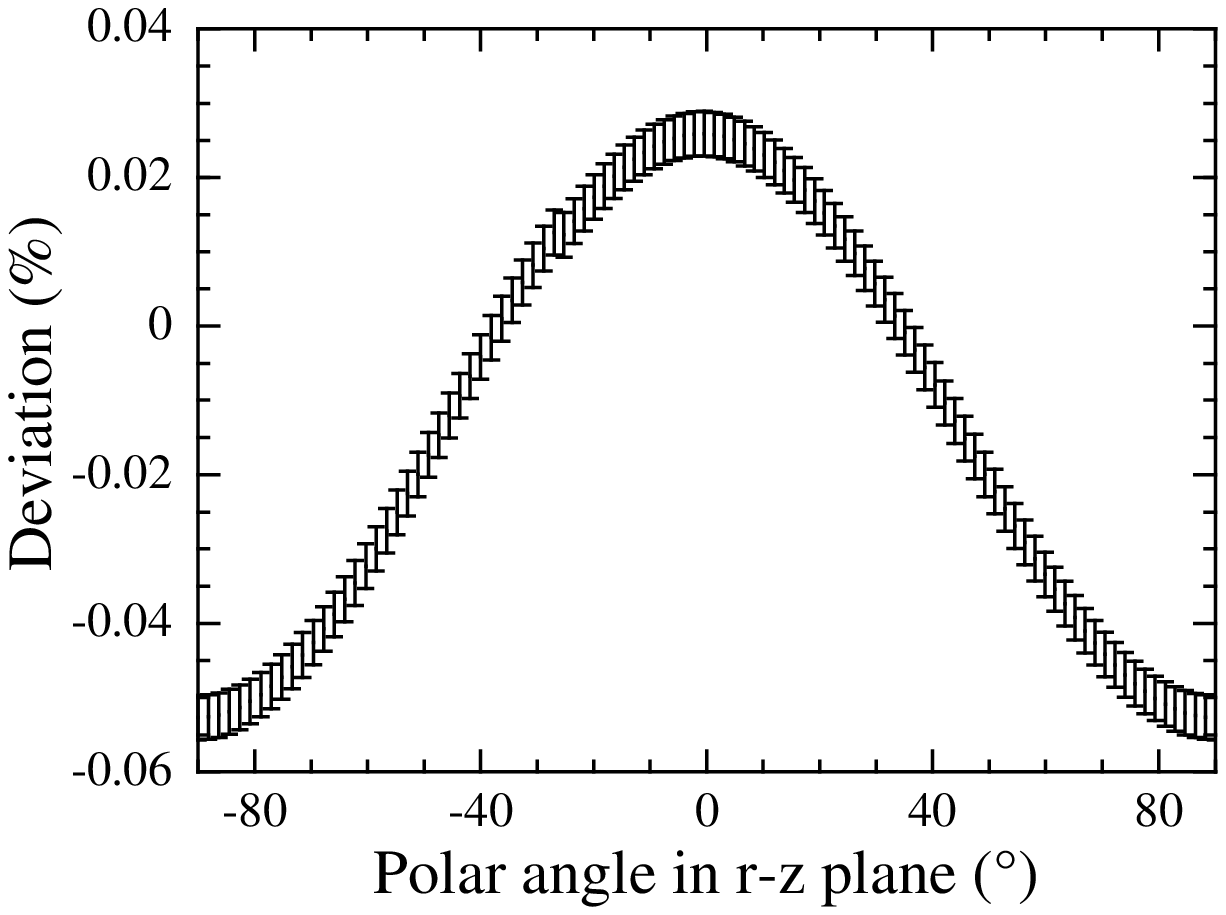}
  \end{center}
  \hfil\small{(a)}\hfil\small{(b)}\hfil\caption{(a)The shape of the gas-liquid
  interface at minimum compensation current $I_{min}$. All lengths are scaled by the cell inner
  radius $R=1.5$ mm. (b) Deviation of the shape in Fig.~\ref{compens}a from the spherical
  shape versus polar angle in the $r-z$ plane. Zero polar angle corresponds to
  the horizontal direction. The error bars show the numerical accuracy of the
  calculations.}\label{compens}
\end{figure}

The bubble shape calculated for the minimum compensation current $I_{min}$ is presented in
Figs.~\ref{compens}.
Only right half of the interface Fig.~\ref{compens}a is calculated. The symmetric left half is added for
illustration purposes only to this and all other figures that show the interface shape. The interface
resembles sphere very much, which explains the experimental result in Fig.~\ref{Sph_field}d. However, there
is a minor deviation from the sphericity, which is shown in Fig.~\ref{compens}b. In the case when the
centers of the shell and of the interface coincide (it is easy to adjust experimentally by moving the
shell with respect to the solenoid), this diagram gives the angular variation of the liquid layer
thickness. Fig.~\ref{compens}b demonstrates that the liquid layer smoothness is at least one order better
than 1\% required for the ICF targets.

Because of the very small value of the van der Waals contribution (indeed, $C_w\sim 10^{-12}$), our code
did not show any effect of the van der Waals forces within the numerical accuracy. The van der Waals
forces should manifest themselves only when the gas bubble is pressed against the shell. Due to them, a
thin wetting layer (layer of liquid phase that separates the gas phase from the shell) forms even in this
situation because of the complete wetting conditions. However, as the thickness of this layer is 100 to
1000 times less than the average thickness in our case, the chosen algorithm does not detect it and
becomes unstable. Therefore, in the following we neglect the presence of the wetting layer, assuming that
its thickness is zero, i.e. the liquid layer is assumed to be discontinuous, the boundary conditions of
the second type (see section \ref{sec3}) being used for this case. We equally neglect the presence of the
van der Waals forces for the case of the continuous liquid layer. The boundary conditions of the first
type are applied in this case.

\subsection{Levitated bubble: effect of the surface tension}

\begin{figure}[htb]
  \begin{center}
  \includegraphics[height=5cm]{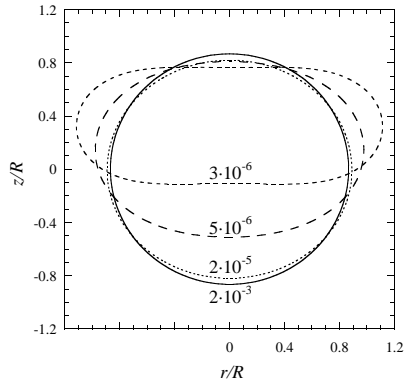}
  \end{center}
  \caption{ The dependence of the shape of the gas bubble levitated in the infinite liquid on the surface tension
  $\sigma$, which is the parameter of the curves expressed in N/m. All shapes are calculated at
  minimum compensation current $I_{min}$. The bubble volume is fixed and corresponds to that of Fig.~\ref{compens}a.}
  \label{suspendu_sigma}
\end{figure}

In this subsection we consider the interface change under the influence of the decrease of the surface
tension. Our experiments carried out for the temperatures close to the critical temperature of the
hydrogen $T_c\approx 33$~K showed a wavy deformation of the gas-liquid interface. In this subsection we
model the hydrogen bubble shape by neglecting the presence of the shell, i.e. by assuming the levitation
of the gas bubble in the large volume of liquid. The results of the calculations are shown in
Fig.~\ref{suspendu_sigma}.
One can see that the bubble deformation increases when the surface tension decreases to zero, the smallest
value of the surface tension corresponding to the temperature of 30 mK below $T_c$. This deformation
appears as a consequence of the residual acceleration $\vec{\gamma}$ inside the solenoid. Qualitatively,
the deformation of the interface corresponds to the observations.

\subsection{Levitated bubble: effect of solenoid current}

\begin{figure}[htb]
  \begin{center}
  \includegraphics[height=5cm]{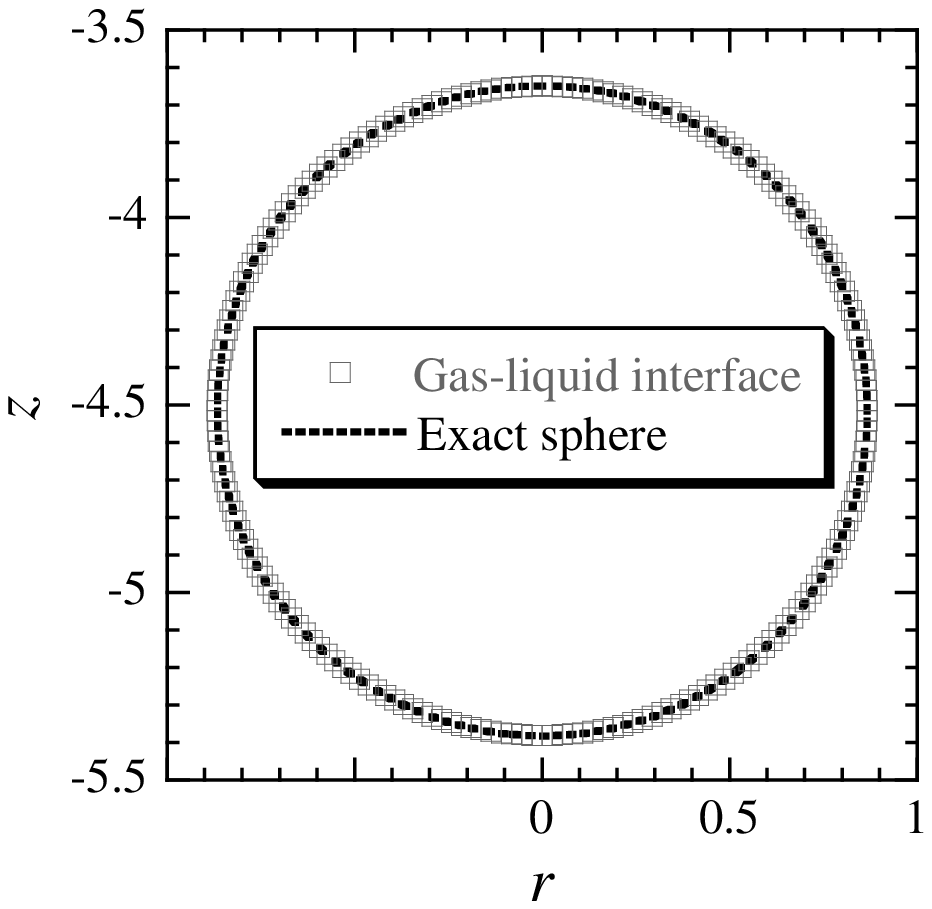}\hspace*{7mm}\includegraphics[height=5cm]{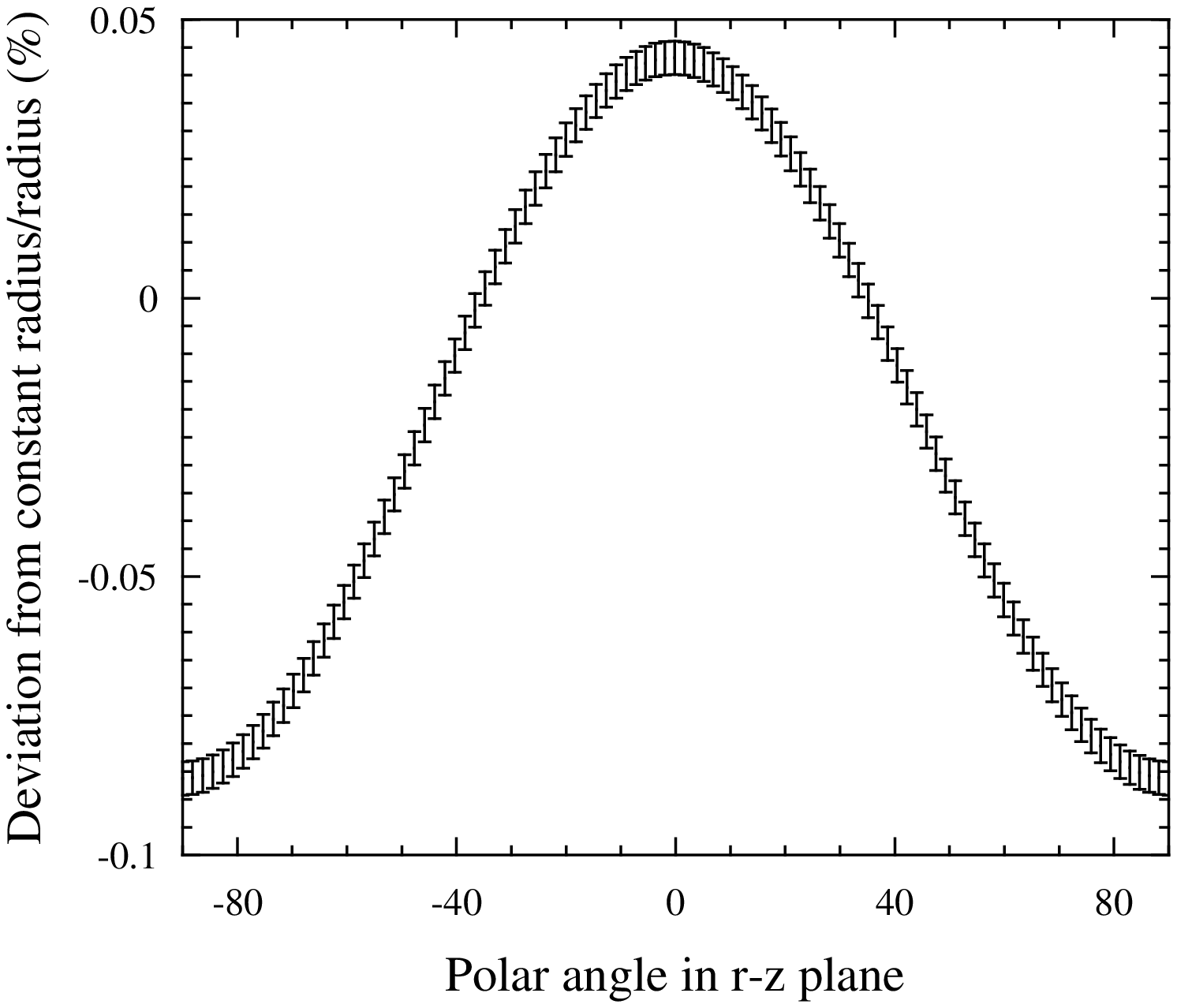}
  \end{center}
\hfil\small{(a)}\hfil\small{(b)}\hfil\caption{(a)The shape of the gas bubble at the current value
$I_{min}\cdot 1.01$. The other values of the
  parameters are the same as for Figs.~\ref{compens}. (b) Deviation of the shape in Fig.~\ref{compens+1}a from the spherical
  shape versus polar angle in the $r-z$ plane.}\label{compens+1}
\end{figure}

The bubble shape calculated for the current one per cent larger than $I_{min}$ is presented in
Figs.~\ref{compens+1}.
Comparing these figures with Figs.~\ref{compens}, two main differences become apparent. First, the
equilibrium position of the bubble is displaced strongly to the bottom of the solenoid, its center
corresponding approximately to the point $z_1$ from Fig.~\ref{force}. In other words, the position of the
levitation point is very sensitive to the value of the current. Second, the maximum deviation from the
average layer thickness is very sensitive to the current. Fig.~\ref{compens+1}b shows that the 1\%
increase of the current causes the $\sim 50$\% increase in the maximum deviation.

For the current less than $I_{min}$, there is no equilibrium levitation position and the bubble is squeezed
against the upper part of the shell.

\subsection{Squeezed bubble}

\begin{figure}[htb]
  \begin{center}
  \includegraphics[height=5cm]{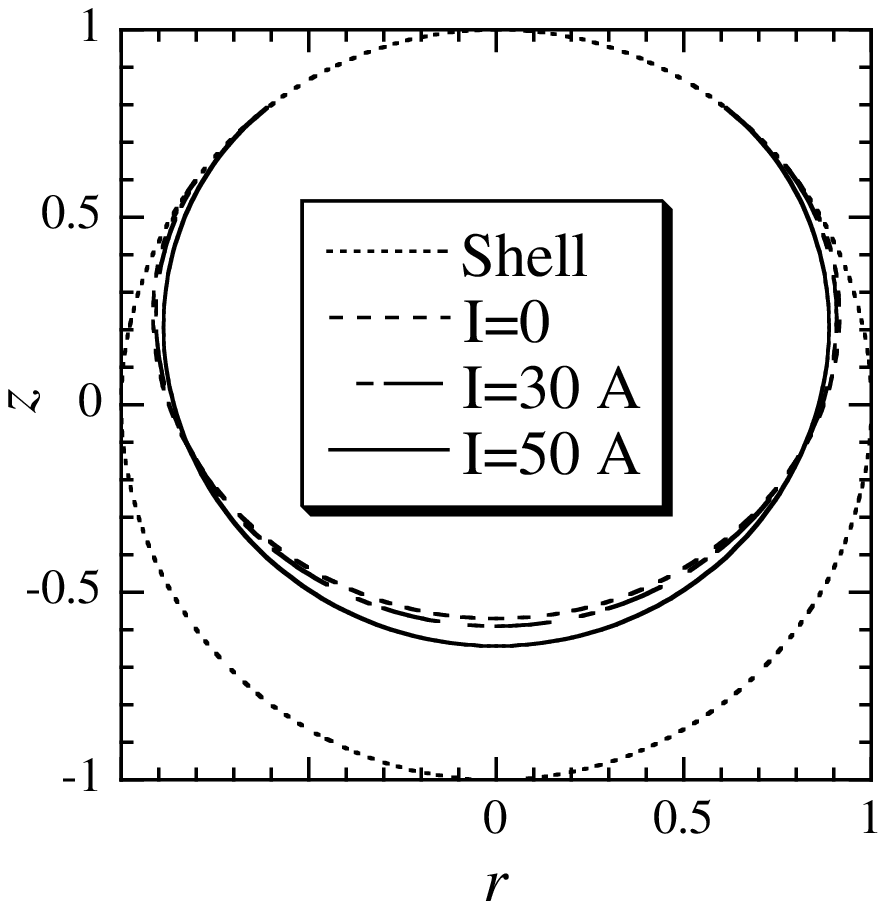}\hspace*{7mm}\includegraphics[height=5cm]{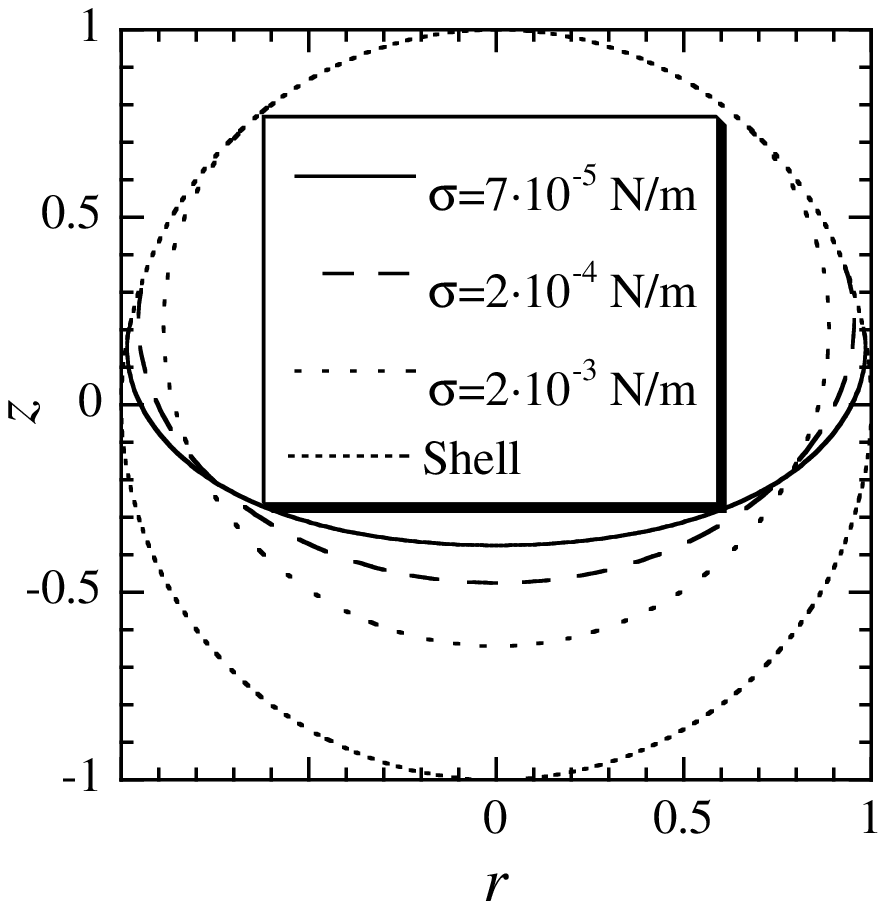}
  \end{center}
  \hfil\small{(a)}\hfil\small{(b)}\hfil\caption{The shapes of the gas bubble pressed against the
  shell calculated for: (a) different values of current and $\sigma=0.002$ N/m; (b) different values of
  $\sigma$ and current $I=50$ A. The other values of the parameters are the same as for
  Figs.~\ref{compens}.}\label{squeezed}
\end{figure}

The squeezed bubble shapes calculated for current values smaller than $I_{min}=60$~A are presented in
Figs.~\ref{squeezed}.
One can see that the variation of the current does not influence the shape strongly at a large value of the
surface tension (Fig.~\ref{squeezed}a). These data compares well to the experiments. However, when the
surface tension decreases, the inhomogeneity of the magnetic forces deform the bubble
(Fig.~\ref{squeezed}b) in full analogy to the case of the levitated bubble shown in
Fig.~\ref{suspendu_sigma}.

\section{Conclusion and perspectives}

The experiments of magnetic levitation of hydrogen, near the triple point, in a hollow semi-transparent
spherical shell have been performed. Due to the fact that the sphere was not completely transparent, a
precise characterization of the liquid layer could not be done. Nevertheless, the obtained images did not
show deformation of the gas/liquid interface under the influence of the residual magnetic forces. Our
numerical modeling of the gas-liquid interface shape shows that the layer of the liquid hydrogen with a
very small thickness variation (less than 0.1\%) can be achieved by means of magnetic levitation in a
superconductive solenoid. The thickness homogeneity is the best when levitating at the minimum value of
the current at which the levitation is possible at all. Although the levitation is possible at a higher
value of the current, the thickness variation increases rapidly with the increase of the current. The
higher homogeneity can be achieved for the a larger surface tension (smaller temperature) and a larger
layer thickness provided that the gas bubble is well centered in the shell.

We investigate now the levitation in a coil based on the magnetic multipole concept. This coil is similar
to coils used for the confinement of particles in accelerators. Such multipoles can levitate hydrogen or
deuterium targets by tens or even hundreds at the same time. This could be a solution to the problem of
high production rates required for commercial power plants based on the inertial fusion energy.

\section*{Acknowledgments}

This work was supported in part by EURATOM. We thank Dr A. Dael of the laboratory of magnetism at
CEA/Saclay and  L. Quettier of the GREEN laboratory at Nancy for their advises on magnetostatics. We thank
D. Beysens for the helpful discussions.

\appendix
\section{Appendix}

Although the final expression for the magnetic contribution (\ref{dpm}) can be found in the literature
\cite{weil}, we never met either its derivation or a discussion of the validity criterium. We address these
two points in this appendix.

The absolute value for the normal component of the force per unit area $p_m$ induced on an interface by
the magnetic field (magnetic pressure) can be calculated in terms of the Cartesian components of the
Maxwell tensor \cite{Land}
\begin{equation}\label{maxw}
  \sigma_{ik}=-{1\over 2}\mu\mu_0H^2\delta_{ik}+\mu\mu_0H_iH_k,
\end{equation}
where $\mu=1+\chi$ is the magnetic permeability of the medium, $H_i$ are the Cartesian components of the
magnetic field, and $\delta_{ik}$ is the Kroneker symbol ($\delta_{ik}=1$ if $i=k$ and 0 otherwise). The
magnetic pressure is then
\begin{equation}\label{pm}
  p_m=\sum\limits_{i,k=1}^3\sigma_{ik}n_in_k={1\over 2}\mu\mu_0(H_n^2-H_\tau^2),
\end{equation}
where $n_i$ are the components of the external normal to the interface, $H_n$ and $H_\tau$ are the normal
and tangential components of the magnetic field, $H^2=H_n^2+H_\tau^2$. Using the boundary conditions
\cite{Land} for the magnetic field at the interface
\begin{equation}\label{bH}
\left\{
  \begin{array}{l}
    \mu_LH_{n L}=\mu_GH_{n G}, \\
    H_{\tau L}=H_{\tau G},
  \end{array}
  \right.
\end{equation}
where the indices L and G refer to the liquid and gas phases, one can obtain the expression
\begin{eqnarray}
  \Delta p_m=p_{mL}-p_{mG}=-{1\over 2}\,\mu_0(\mu_L-\mu_G)\left(H_{\tau G}^2+
  {\mu_G\over\mu_L}H_{n G}^2\right)\approx\nonumber\\
-{1\over 2}\,\mu_0(\mu_L-\mu_G)H_G^2\approx-{1\over 2\mu_0}\,(\mu_L-\mu_G)B^2,\label{dpm}
\end{eqnarray}
where two approximations were used: $\mu_L\approx\mu_G\approx 1$. These approximate equalities verify with
the accuracy $10^{-6}$ in our case and thus the final expression (\ref{dpm}) can be employed.

The volume magnetic force (\ref{dfm}) can be obtained from (\ref{dpm}) using the integral Gauss theorem
\cite{Land} as $\Delta \vec{f}_m=\nabla(\Delta p_m)$. By converting $\Delta p_m$ into the a non-dimensional
form, one obtains the final expression
\begin{equation}\label{dpmn}
  \Delta p_m=\mathrm{Mg}\, B(r,z)^2,
\end{equation}
where $B$ is expressed in units of $B(0,0)$, and
\begin{equation}\label{mg}
  \mathrm{Mg}={(\rho_L-\rho_G)|\alpha|R\over 2\sigma\mu_0}\,B^2(0,0)
\end{equation}
is the non-dimensional number that characterizes the strength of the magnetic force.

\end{document}